\documentclass[conference]{IEEEtran}
\IEEEoverridecommandlockouts
\usepackage{hyperref}
\usepackage{cite}
\usepackage{amsmath,amssymb,amsfonts}
\usepackage{algorithmic}
\usepackage{graphicx}
\usepackage{textcomp}
\usepackage{cite}
\usepackage{array}
\usepackage[table,xcdraw]{xcolor}
\usepackage[numbers]{natbib}
\usepackage{booktabs}
\usepackage{lipsum}
\usepackage{multicol}
\usepackage{etoolbox}
\usepackage{relsize}
\def\BibTeX{{\rm B\kern-.05em{\sc i\kern-.025em b}\kern-.08em
    T\kern-.1667em\lower.7ex\hbox{E}\kern-.125emX}}

\makeatletter
\patchcmd{\@IEEEauthorblockA}{\normalfont\normalsize}{\normalfont}{}{}
\makeatother

\makeatletter
\patchcmd{\@IEEEauthorblockA}{\normalfont\normalsize}{\normalfont}{}{}
\makeatother

\makeatletter

\def\ps@IEEEtitlepagestyle{%
  \def\@oddfoot{\mycopyrightnotice}%
  \def\@evenfoot{}%
}
\def\mycopyrightnotice{%
  {\footnotesize 979-8-3503-9114-5/24/\$31.00~\copyright~2024 IEEE\hfill}
  \gdef\mycopyrightnotice{}
}

\begin{document}

\title{Thyroidiomics: An Automated Pipeline for Segmentation and Classification of Thyroid Pathologies from Scintigraphy Images\\
{\footnotesize}
\thanks{\textsuperscript{*} Equal contribution}
}
\author{
\IEEEauthorblockN{Maziar Sabouri*}
\IEEEauthorblockA{ \textit{University of British Columbia}\\
 Vancouver, Canada \\
 maziarsabouri7@gmail.com}
\and
\IEEEauthorblockN{Shadab Ahamed*}
\IEEEauthorblockA{ \textit{University of British Columbia}\\
 Vancouver, Canada \\
 shadabahamed1996@gmail.com}
\and
\IEEEauthorblockN{Azin Asadzadeh}
\IEEEauthorblockA{ \textit{Golestan University of Medical Sciences}\\
 Gorgan, Iran}
\and
\IEEEauthorblockN{Atlas Haddadi Avval}
\IEEEauthorblockA{ \textit{Mashhad University of Medical Sciences}\\
 Mashhad, Iran}
\and
\IEEEauthorblockN{Soroush Bagheri}
\IEEEauthorblockA{ \textit{Kashan University of Medical Sciences}\\
 Kashan, Iran}
\and
\IEEEauthorblockN{Mohsen Arabi}
\IEEEauthorblockA{ \textit{Alborz University of Medical Sciences}\\
 Karaj, Iran}
\and
\IEEEauthorblockN{Seyed Rasoul Zakavi}
\IEEEauthorblockA{ \textit{Mashhad University of Medical Sciences}\\
 Mashhad, Iran}
\and
\IEEEauthorblockN{Emran Askari}
\IEEEauthorblockA{ \textit{Mashhad University of Medical Sciences}\\
 Mashhad, Iran}
\and
\IEEEauthorblockN{Ali Rasouli}
\IEEEauthorblockA{ \textit{Kashan University of Medical Sciences}\\
 Kashan, Iran}
\and
\IEEEauthorblockN{Atena Aghaee}
\IEEEauthorblockA{ \textit{Mashhad University of Medical Sciences}\\
 Mashhad, Iran}
\and
\IEEEauthorblockN{Mohaddese Sehati}
\IEEEauthorblockA{ \textit{Golestan University of Medical Sciences}\\
 Gorgan, Iran}
\and
\IEEEauthorblockN{Fereshteh Yousefirizi}
\IEEEauthorblockA{ \textit{BC Cancer Research Institute}\\
 Vancouver, Canada}
\and
\IEEEauthorblockN{Carlos Uribe}
\IEEEauthorblockA{ \textit{BC Cancer Research Institute}\\
 Vancouver, Canada}
\and
\IEEEauthorblockN{Ghasem Hajianfar}
\IEEEauthorblockA{ \textit{Geneva University Hospital}\\
 Geneva, Switzerland}
\and
\IEEEauthorblockN{Habib Zaidi}
\IEEEauthorblockA{ \textit{Geneva University Hospital}\\
 Geneva, Switzerland}
\and
\IEEEauthorblockN{Arman Rahmim}
\IEEEauthorblockA{ \textit{University of British Columbia}\\
 Vancouver, Canada }
}

\maketitle

\begin{abstract}
The objective of this study was to develop an automated pipeline that enhances thyroid disease classification using thyroid scintigraphy images, aiming to decrease assessment time and increase diagnostic accuracy. Anterior thyroid scintigraphy images from 2,643 patients were collected and categorized into diffuse goiter (DG), multinodal goiter (MNG), and thyroiditis (TH) based on clinical reports, and then segmented by an expert. A Residual UNet (ResUNet) model was trained to perform auto-segmentation. Radiomic features were extracted from both physician (scenario 1) and ResUNet segmentations (scenario 2), followed by omitting highly correlated features using Spearman's correlation, and feature selection using Recursive Feature Elimination (RFE) with eXtreme Gradient Boosting (XGBoost) as the core. All models were trained under leave-one-center-out cross-validation (LOCOCV) scheme, where nine instances of algorithms were iteratively trained and validated on data from eight centers and tested on the ninth for both scenarios separately. Segmentation performance was assessed using the Dice similarity coefficient (DSC), while classification performance was assessed using metrics, such as precision, recall, F1-score, accuracy, area under the Receiver Operating Characteristic (ROC AUC), and area under the precision-recall curve (PRC AUC). ResUNet achieved DSC values of 0.84$\pm$0.03, 0.71$\pm$0.06, and 0.86$\pm$0.02 for MNG, TH, and DG, respectively. Classification in scenario 1 achieved an accuracy of 0.76$\pm$0.04 and a ROC AUC of 0.92$\pm$0.02 while in scenario 2, classification yielded an accuracy of 0.74$\pm$0.05 and a ROC AUC of 0.90$\pm$0.02. The automated pipeline demonstrated comparable performance to physician segmentations on several classification metrics across different classes, effectively reducing assessment time while maintaining high diagnostic accuracy. The code is available at:
\href{https://github.com/ahxmeds/thyroidiomics.git}{\textcolor{blue}{\textit{https://github.com/ahxmeds/thyroidiomics.git}}}.
\end{abstract}


\section{Introduction}
Thyroid diseases are the second most prevalent endocrine diseases, with an estimated number of over 300 million new cases per year worldwide \cite{Parry2017}. It is, therefore, crucial to differentiate thyroid pathologies. Imaging plays an important role in the diagnosis and assessment of these pathologies, including but not limited to planar scintigraphy, ultrasound, single-photon emission computed tomography, and computed tomography scans. In most cases, physicians analyze the images and might gather other demographical, clinical, and laboratory data in order to make a definite diagnosis. However, some pathologies are hard to differentiate based on the imaging data and subtle differences are not evident to the naked eye. In addition,  assessment of images might be time and labor-intensive, and prone to errors due to expert subjectivity. Therefore, it is of utmost importance to develop solutions to facilitate automated thyroid disease classification. Meanwhile, radiomics analyses - involving quantitative features extracted from regions of interest (ROIs) - combined with artificial intelligence (AI), including machine learning (ML) and deep learning (DL), have shown significant potential and value as applied to medical images \cite{sabouri2023myocardial, shiri2021machine}. For thyroid diseases, AI is a promising solution that enhances the accuracy of thyroid pathology detection and classification in various imaging modalities. \cite{Currie2021, ma2019thyroid, wei2020ensemble, zhang2024novel, yu2017computer, zhang2022deep, li2022contrast}. 

Several existing studies have implemented AI solutions for $^\text{99m}$Technetium Pertechnetate (99m Tc) thyroid scintigraphy scans. In \cite{Currie2021}, ML and DL models were applied to thyroid scans to predict the type of disease, hyperthyroidism or hypothyroidism. Their model achieved a classification accuracy of 0.80 based on images only. Thyroid uptake pattern recognition network (TPRNet) was proposed in \cite{pi2022fusing} to identify the focal versus diffuse change patterns of thyroid scintigraphy scans. Their patient cohort was large ($>$5000 scans) and with the trained model’s classification performance comparable to that by physician visual assessment. In \cite{Ma2019-xc}, a modified DenseNet was employed to classify over 2800 scintigraphic images from a single center. Experimental results indicate that the model is highly effective for diagnosing thyroid diseases using scintigraphic images, outperforming other CNN methods. Another work in \cite{Zhao2023-wg} improved disease diagnosis by utilizing a ResNet-34 based network to classify thyroid scintigraphic images using a dataset with around 3200 images. Another diagnostic study was performed via transfer learning, utilizing pre-trained AlexNet, VGGNet, and ResNet to classify thyroid pathologies \cite{qiao2021deep}. The accuracy and F1-score were compared with nuclear medicine residents, showing that all models outperformed first-year residents, but had suboptimal performance as compared to third-year residents (model metrics values ranged from 0.73 to 0.97). Another study \cite{yang2021automatic} classified the thyroid uptake patterns via transfer learning, showing that the InceptionV3 network reached the highest performance (accuracy of 0.88 in the external set).To the best of our knowledge, no study exists that assesses the use of ML or DL methods to predict the final thyroid pathology such as diffuse goiter (DG), multinodular goiter (MNG), and thyroiditis (TH). 

In this work, we propose an automated pipeline, branded as \textit{Thyroidiomics}, whos input is a thyroid scintigraphy image and outputs the final pathology class. In addition, we compared the results of our proposed model with the ground truth segmentation-derived radiomics classification.

\section{Material and Methods}

\begin{figure}[!t]
\centering
\includegraphics[width=0.95\linewidth]{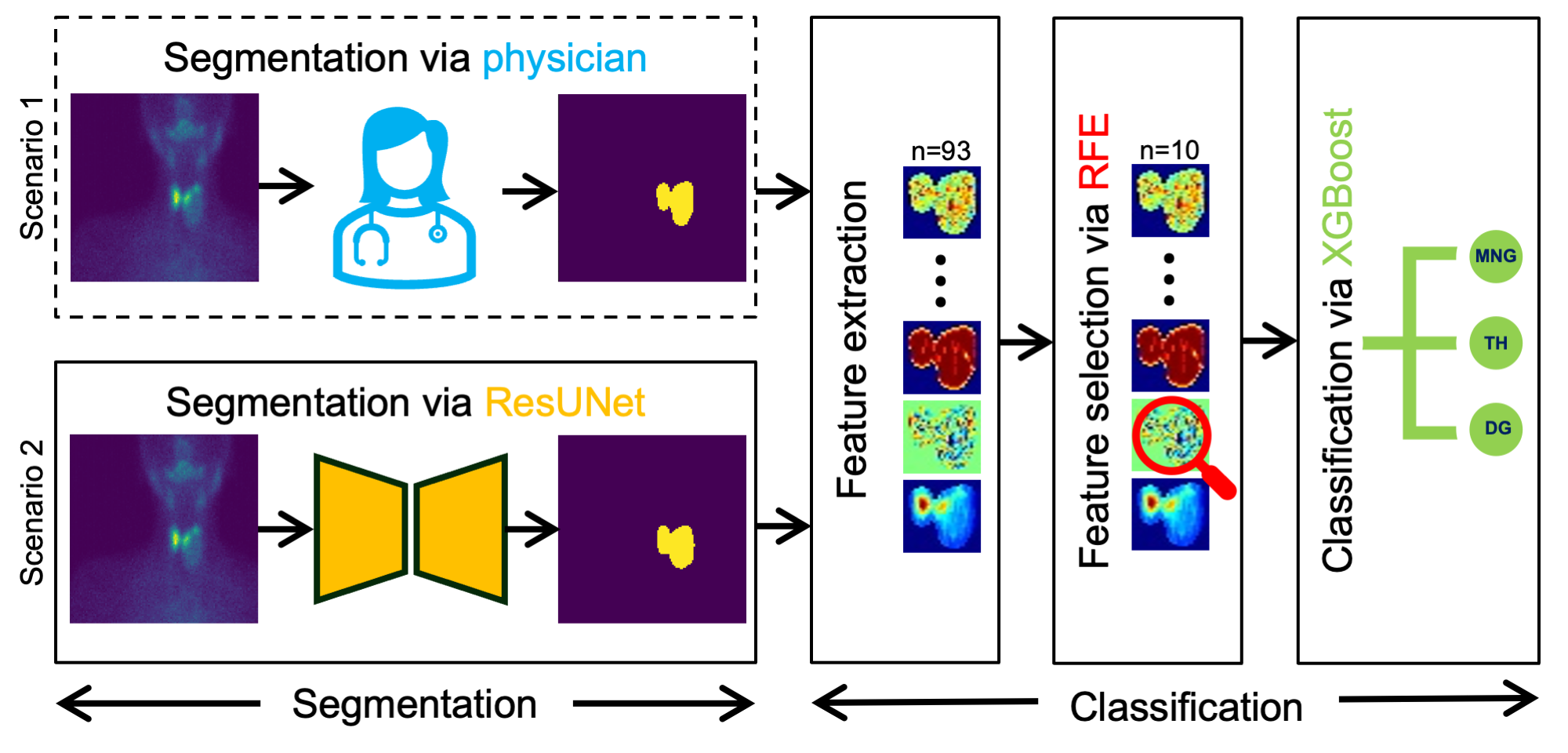}
\caption{\textit{Thyroidiomics}: the proposed two-step pipeline to classify thyroid pathologies into three classes, namely, MNG, TH and DG. Scenario 1 represents the pipeline dependent on physician delineated ROIs as input to the classifier, while scenario 2 represents the fully automated pipeline operating on segmentation predicted by ResUNet.}
\label{Fig: Flowchart}
\end{figure}

Fig. \ref{Fig: Flowchart} summarizes our proposed automated pipeline that uses a two-step process for thyroid pathology classification: 
\begin{enumerate}
\item Thyroid segmentation: a trained Residual UNet (ResUNet) is used to segment the thyroid region from scintigraphy images, obtaining an ROI for input to the second step,
\item Pathology classification: radiomics features are extracted from the segmented ROIs generated by ResUNet which are then fed to a trained classifier to predict the final pathology class. 
\end{enumerate}

\subsection{Data collection and preprocessing}
In this work, we used anterior thyroid scintigraphy images from 2,643 patients collected at nine different medical centers, categorized into three different pathology classes encompassing MNG, TH, and DG based on the corresponding clinical reports. Furthermore, the whole thyroid region was segmented from the scintigraphy images by an experienced nuclear medicine physician using the manual contouring tool from the ITK-Snap software \cite{py06nimg}. Table \ref{tab:data_table} summarizes the data on age, sex, number of cases in each pathology, and scanner manufacturers across different centers. Approval for the use of data in this study was granted by the research ethics board of each respective institution.

Out of the 2,643 images, 278 (all belonging to center 5) were images of size \(256 \times 256\), while the remaining were of size \(128 \times 128\). The images from center 5 were resampled to \(128 \times 128\) using nearest neighbor interpolation.

\begin{table}[]
\centering
\caption{Demographics, manufacturer distribution, and pathology frequencies of patients across centers. The age information from center 1 was not available. Here, MNG: multinodular goiter, TH: thyroiditis, DG: diffuse goiter, F: Female, and M: Male.}
\label{tab:data_table}
\resizebox{0.98\columnwidth}{!}{%
\huge
\begin{tabular}{ccccc}
\hline
\textbf{Center} &
  \textbf{\begin{tabular}[c]{@{}c@{}}Median age \\ {[}Inter-quartile \\ range{]} \\ (years)\end{tabular}} &
  \textbf{\begin{tabular}[c]{@{}c@{}}Sex \\ (F / M)\end{tabular}} &
  \textbf{\begin{tabular}[c]{@{}c@{}}Classes \\ (MNG / TH / DG)\end{tabular}} &
  \textbf{Vendor} \\ \hline
1 &
  - &
  281 / 105 &
  212 / 77 / 97 &
  \begin{tabular}[c]{@{}c@{}}ADAC\\ GENESYS\end{tabular} \\ \hline
2 &
  \begin{tabular}[c]{@{}c@{}}43\\{[}36, 56{]}\end{tabular} &
  155 / 67 &
  116 / 44 / 62 &
  \begin{tabular}[c]{@{}c@{}}Siemens NM \\IP2 (ECAM1028)\end{tabular} \\ \hline
3 &
  \begin{tabular}[c]{@{}c@{}}44\\{[}35, 56{]}\end{tabular} &
  496 / 130 &
  333 / 80 / 213 &
  \begin{tabular}[c]{@{}c@{}}Siemens NM \\IP2 (ECAM1028)\end{tabular} \\ \hline
4 &
   \begin{tabular}[c]{@{}c@{}}42\\{[}35, 54{]}\end{tabular} &
  360 / 116 &
  202 / 117 / 157 &
  \begin{tabular}[c]{@{}c@{}}Mediso\\ AnyScanS\end{tabular} \\ \hline
5 &
   \begin{tabular}[c]{@{}c@{}}47\\{[}37, 59{]}\end{tabular} &
  211 / 67 &
  139 / 57 / 82 &
  \begin{tabular}[c]{@{}c@{}}Siemens NM \\IP1 (ECAM10482)\end{tabular} \\ \hline
6 &
   \begin{tabular}[c]{@{}c@{}}39\\{[}36, 52{]}\end{tabular} &
  47 / 18 &
  21 / 23 / 21 &
  MiE SCINTRON \\ \hline
7 &
   \begin{tabular}[c]{@{}c@{}}41\\{[}31, 51{]}\end{tabular} &
  237 / 89 &
  76 / 50 / 200 &
  \begin{tabular}[c]{@{}c@{}}GE Discovery \\ NM 630\end{tabular} \\ \hline
8 &
   \begin{tabular}[c]{@{}c@{}}47\\{[}37, 60{]}\end{tabular} &
  101 / 44 &
  55 / 26 / 64 &
  \begin{tabular}[c]{@{}c@{}}Siemens Encore 2\\ Symbia 1071\end{tabular} \\ \hline
9 &
   \begin{tabular}[c]{@{}c@{}}41\\{[}32, 51{]}\end{tabular} &
  89 / 30 &
  29 / 41 / 49 &
  GE INFINIA \\ \hline
\end{tabular}%
}
\end{table}

\subsection{Experimental design: Leave-one-center-out}
In this study, we employed leave-one-center-out cross-validation (LOCOCV) for training and testing our models \cite{yousefirizi2021segmentation}. For data from nine centers, this method involved training a model on data from all centers except one, which was used as the test set, thereby giving us a set of nine trained models. LOCOCV helps reduce the risk of bias towards data from any particular center and helps estimate the robustness of the pipeline. The performance metrics from all iterations are aggregated to obtain a comprehensive assessment of the model’s overall performance and generalizability to data from unseen centers. Both segmentation and classification models were trained under the LOCOCV scheme independently, supervised by their respective ground truth labels. 

\subsection{Segmentation}
\begin{enumerate}
    \item \textit{Preprocessing and augmentation}: All input image intensities were clipped in the range of (0, 550), and normalized to (0,1) thereafter. Both input and mask were then resampled to pixel spacing of 1 mm $\times$ 1 mm via bilinear and nearest-neighbor interpolations, respectively. Finally, we employed a series of randomized transforms: (i) patches of size 64$\times$64 were cropped from both input and mask images with the center of the patch lying on a thyroid pixel with a probability of 2/3, while lying on a background pixel with a probability of 1/3; (ii) random 2D translation in range (-5,5) pixels; (iii) random rotation in ($-\pi/12, \pi/12$); (iv) random scaling by 1.1 in 2D. A different set of randomized transforms was called at the start of every epoch for all images to improve the diversity of images seen by the model during training.
    \item \textit{Network}: We used a ResUNet \cite{resunet, monai} consisting of 5 layers of encoder and decoder (with residual blocks) paths with skip-connections. The encoders in each layer consisted 16, 32, 64, 128, 256 convolutional filters, respectively. The input in the encoder was downsampled using strided convolutions (stride=2), while the decoder unsampled using transpose strided convolutions. PReLU was used as activation function throughout the network. This network consisted of 1,627,597 trainable parameters.
    \item \textit{Loss function, optimizer, learning rate scheduler}: We utilized a Dice loss with a weighted (using $\alpha$) regularization term for penalizing the prediction of false positives. The final loss function $\mathcal{L}$ is given by,
    \begin{equation}
        \mathcal{L} = 1 - \frac{2\sum_i p_i g_i + \epsilon}{\sum_i p_i + \sum_i g_i + \epsilon} + \alpha \cdot \frac{\sum_i p_i(1-g_i)}{\sum_i p_i + \sum_i g_i + \epsilon}
    \end{equation}
    where, $\sum_i$ denotes a sum over all pixels in the patch, $p_i$ denotes the predicted value of the class (0 or 1) (after applying softmax) and $g_i$ denotes the GT mask value at pixel $i$. In our experiments, we set $\alpha = 2$. Small constants $\epsilon$ were added in the numerator and denominator for numerical stability of the loss function during training. The loss optimized using AdamW optimizer with an initial learning rate $2\times10^{-4}$ and weight-decay of $1\times10^{-5}$. The learning rate was modulated from the initial value to zero (in 300 epochs) using a cosine annealing scheduler. The model with the highest Dice similarity coefficient (DSC) averaged over all batches (batch size = 32) on the validation set was chosen as the best model over 300 epochs.  
    \item \textit{Sliding window inference}: The same intensity clipping/normalization and resampling were applied on test images as before. The inference was performed on the test set using the sliding window approach with a window size of 128$\times$128. The test set predictions were resampled to the coordinates of the original GT masks for computing the evaluation metrics. 
    \item \textit{Evaluation metrics}: On the test set, the models were evaluated using the DSC metric. 
\end{enumerate}

\subsection{Classification}
\begin{enumerate}
\item \textit{Feature extraction}: We performed feature extraction using the Pyradiomics library \cite{van2017computational}, aligned with IBSI guidelines \cite{zwanenburg2020image}. First, images were z-score normalized and resampled to a pixel spacing of 1 mm $\times$ 1 mm using BSpline interpolation. Radiomics features were then extracted from the thyroid ROIs with a bin width of 0.3. A total of 93 features were extracted, comprising 18 first-order (FO), 24 gray-level co-occurrence matrix (GLCM), 14 gray-level dependence matrix (GLDM), 16 gray-level run length matrix (GLRLM), 16 gray-level size zone matrix (GLSZM), and 5 neighboring gray-tone difference matrix (NGTDM) features. Features were extracted both from physician delineated ROIs (scenario 1) as well as ResUNet predictions (scenario 2). Classification was performed separately for each scenario, using the same preprocessing steps, feature selection method, and machine learning model discussed below. 
\item \textit{Preprocessing, feature selection, and classification algorithm}: The 93 extracted features were z-score normalized. We used Spearman's correlation to identify and remove highly correlated features (with correlation coefficient $>$ 0.95). Subsequently, we used the recursive feature elimination (RFE) method with an eXtreme Gradient Boosting (XGBoost) classifier as the core to select 10 features. The training process utilized the XGBoost classifier with 5-fold cross-validation to select the best hyperparameters via grid search. The best model was assessed on two separate test sets: one comprising features extracted from the physician delineated ROIs and another from ResUNet ROI predictions from the same center (under the LOCOCV scheme).
\item \textit{Evaluation metrics}: On the test set, the models were evaluated using various classification metrics like precision, recall, F1-score, accuracy, area under the Receiver Operating Characteristic (ROC AUC), and the precision-recall (PRC AUC) curves. Due to the multiclass nature of our problem, we used different averaging techniques like micro, macro, and weighted averaging over various metrics from different classes \cite{grandini2020metrics}.
\end{enumerate}

\section{Results and Discussion}
The results of our method are systematically presented and analyzed across various key sections: (A) Selected features, (B) classification based on features extracted from physician delineated ROIs, (C) performance of segmentation network, and (D) classification based on features extracted from ResUNet predicted ROIs (fully automated pipeline).

\subsection{Selected features}
Table \ref{tab:feature_selection} shows the selected features with their importance and frequency of selection over different LOCOCV training regimes. It also includes feature map plots for one case from each class for visualization and interpretability. FO-Kurtosis, FO-Skewness, GLDM-Dependence non-uniformity normalized (GLDM-DNUN), and GLDM-Dependence non-uniformity (GLDM-DNU) were selected 9 times, followed by NGTDM-Coarseness and NGTDM-Contrast each selected 8 times, and NGTDM-Complexity with 7 selections. This indicates their consistent importance across various centers or types of data, regardless of different acquisition times, radiopharmaceutical activity, zooming, etc. Regarding the average importance values, GLCM-Cluster shade (GLCM-CS), GLDM-Gray level non-uniformity (GLDM-GLNU), and NGTDM-Complexity received the top scores of 0.246, 0.224, and 0.114, respectively.

It should also be noted that among the 90 possible selected features (9 centers $\times$ 10 features), FO and GLDM each had the highest selection frequency with 25 features each. This was followed by NGTDM with 23 features. Finally, GLCM and GLRLM had the lowest selection frequencies with 12 and 5 features, respectively.

\begin{table}[h]
\centering
\caption{Average importance and count of features selected using RFE with XGBoost across all centers. Feature map plots for one case from each class are also included for visualization and interpretability.  (FO: First-order, DNUN: dependence non-uniformity normalized, DNU: dependence non-uniformity, CS: cluster shade, IV: inverse variance, GLNU: gray level non-uniformity, DV: dependence variance).}
\label{tab:feature_selection}
\resizebox{1\columnwidth}{!}{%
\large
\begin{tabular}{>{\centering\arraybackslash}m{2cm}>{\centering\arraybackslash}m{2cm}>{\centering\arraybackslash}m{2cm}>{\centering\arraybackslash}m{2cm}>{\centering\arraybackslash}m{2cm}}
\hline
\textbf{Feature} &
  \textbf{\begin{tabular}[c]{@{}c@{}}Importance \\ average \\(Number\\ of selections)\end{tabular}} &
  \textbf{MNG} &
  \textbf{TH} &
  \textbf{DG} \\ \hline
FO-Kurtosis      & 0.099 (9) & \includegraphics[width=1cm]{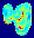} & \includegraphics[width=1cm]{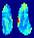} & \includegraphics[width=1cm]{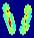} \\
FO-Skewness      & 0.082 (9) & \includegraphics[width=1cm]{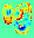} & \includegraphics[width=1cm]{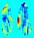} & \includegraphics[width=1cm]{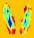} \\
GLDM-DNU         & 0.092 (9) & \includegraphics[width=1cm]{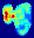} & \includegraphics[width=1cm]{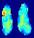} & \includegraphics[width=1cm]{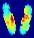} \\
GLDM-DNUN        & 0.061 (9) & \includegraphics[width=1cm]{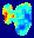} & \includegraphics[width=1cm]{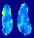} & \includegraphics[width=1cm]{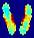} \\
NGTDM-Coarseness & 0.092 (8) & \includegraphics[width=1cm]{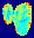} & \includegraphics[width=1cm]{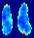} & \includegraphics[width=1cm]{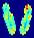} \\
NGTDM-Contrast   & 0.069 (8) & \includegraphics[width=1cm]{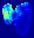} & \includegraphics[width=1cm]{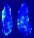} & \includegraphics[width=1cm]{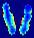} \\
NGTDM-Complexity & 0.114 (7) & \includegraphics[width=1cm]{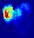} & \includegraphics[width=1cm]{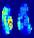} & \includegraphics[width=1cm]{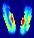} \\
FO-Minimum       & 0.059 (6) & \includegraphics[width=1cm]{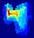} & \includegraphics[width=1cm]{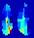} & \includegraphics[width=1cm]{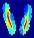} \\
GLCM-CS          & 0.246 (6) & \includegraphics[width=1cm]{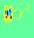} & \includegraphics[width=1cm]{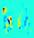} & \includegraphics[width=1cm]{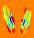} \\
GLCM-IV          & 0.051 (6) & \includegraphics[width=1cm]{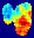} & \includegraphics[width=1cm]{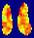} & \includegraphics[width=1cm]{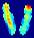} \\
GLDM-GLNU        & 0.224 (6) & \includegraphics[width=1cm]{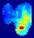} & \includegraphics[width=1cm]{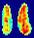} & \includegraphics[width=1cm]{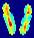} \\
GLRLM-GLNU       & 0.065 (5) & \includegraphics[width=1cm]{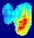} & \includegraphics[width=1cm]{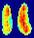} & \includegraphics[width=1cm]{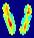} \\
FO-10Percentile  & 0.048 (1) & \includegraphics[width=1cm]{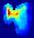} & \includegraphics[width=1cm]{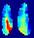} & \includegraphics[width=1cm]{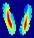} \\
GLDM-DV          & 0.051 (1) & \includegraphics[width=1cm]{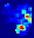} & \includegraphics[width=1cm]{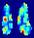} & \includegraphics[width=1cm]{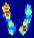} \\ \hline
\end{tabular}%
}
\end{table}

\subsection{Classification using features extracted via physician segmentations}

\begin{figure}[!ht]
\centering
\includegraphics[width=1\columnwidth]{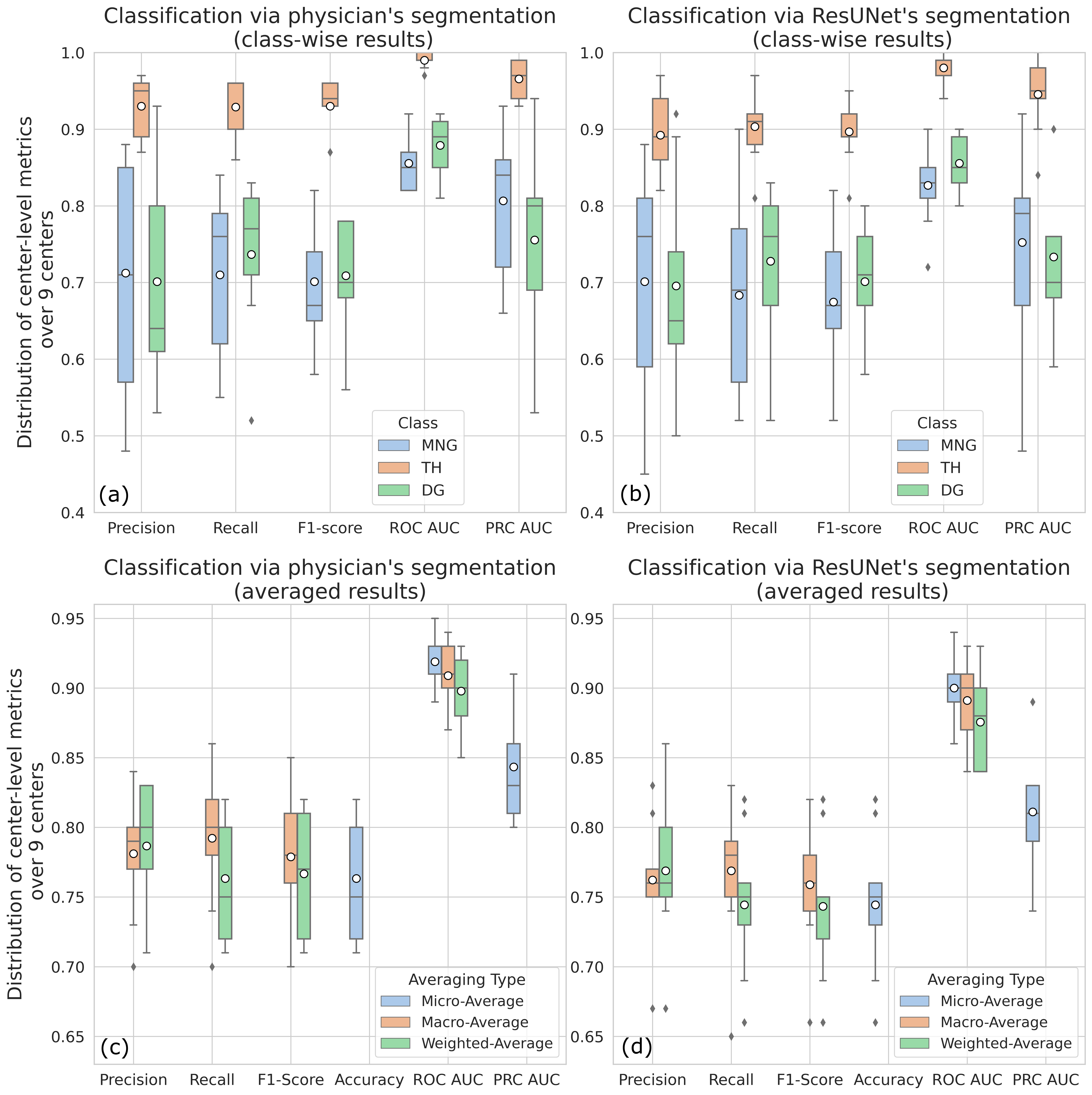} 
\caption{Various class-wise and averaged metrics for classification were used to evaluate model performance in two scenarios: features extracted from the physician delineated ROIs and those from ResUNet predicted ROIs. The boxplots show the distribution of metrics over the nine centers as test sets for the three thyroid pathology classes, MNG, TH and DG. The black horizontal lines denote the median and white circle denote the mean of distribution.}
\label{Fig: Metrics}
\end{figure}

All metrics are visualized as distribution over nine centers using boxplots in Fig. \ref{Fig: Metrics} for both scenarios, encompassing features extracted from physician delineated and ResUNet predicted ROIs. 

As can be seen from Fig. \ref{Fig: Metrics} (c), the micro-averaging yielded a mean accuracy of 0.76$\pm$0.04, ROC AUC of 0.92$\pm$0.02, and PRC AUC of 0.84$\pm$0.04 across all centers, reflecting high consistency. The macro-average approach achieved a mean precision of 0.78$\pm$0.04, recall of 0.79$\pm$0.05, F1-score of 0.79$\pm$0.05, and ROC AUC of 0.91$\pm$0.02, indicating reliable performance. The weighted-average method had a similar performance with a mean precision of 0.79$\pm$0.05, recall of 0.76$\pm$0.04, F1-score of 0.77$\pm$0.04, and ROC AUC of 0.90$\pm$0.03, demonstrating consistent results.

We also analyzed the results for each class individually (Fig. \ref{Fig: Metrics} (a)). The TH class metrics had high performance with a mean precision of 0.93$\pm$0.04, recall of 0.93$\pm$0.04, F1-score of 0.93$\pm$0.04, ROC AUC of 0.99$\pm$0.01, and PRC AUC of 0.97$\pm$0.03 across all centers, suggesting consistent and reliable results. The MNG class showed moderate performance with a mean precision of 0.71$\pm$0.15, recall of 0.71$\pm$0.10, F1-score of 0.70$\pm$0.08, ROC AUC of 0.86$\pm$0.04, and PRC AUC of 0.81$\pm$0.09. The DG class had performance comparable to MNG class with a mean precision of 0.70$\pm$0.13, recall of 0.74$\pm$0.10, F1-score of 0.71$\pm$0.08, ROC AUC of 0.88$\pm$0.04, and PRC AUC of 0.76$\pm$0.12.

\subsection{ResUNet segmentation}
\begin{figure}[!ht]
\centering
\includegraphics[width=\columnwidth]{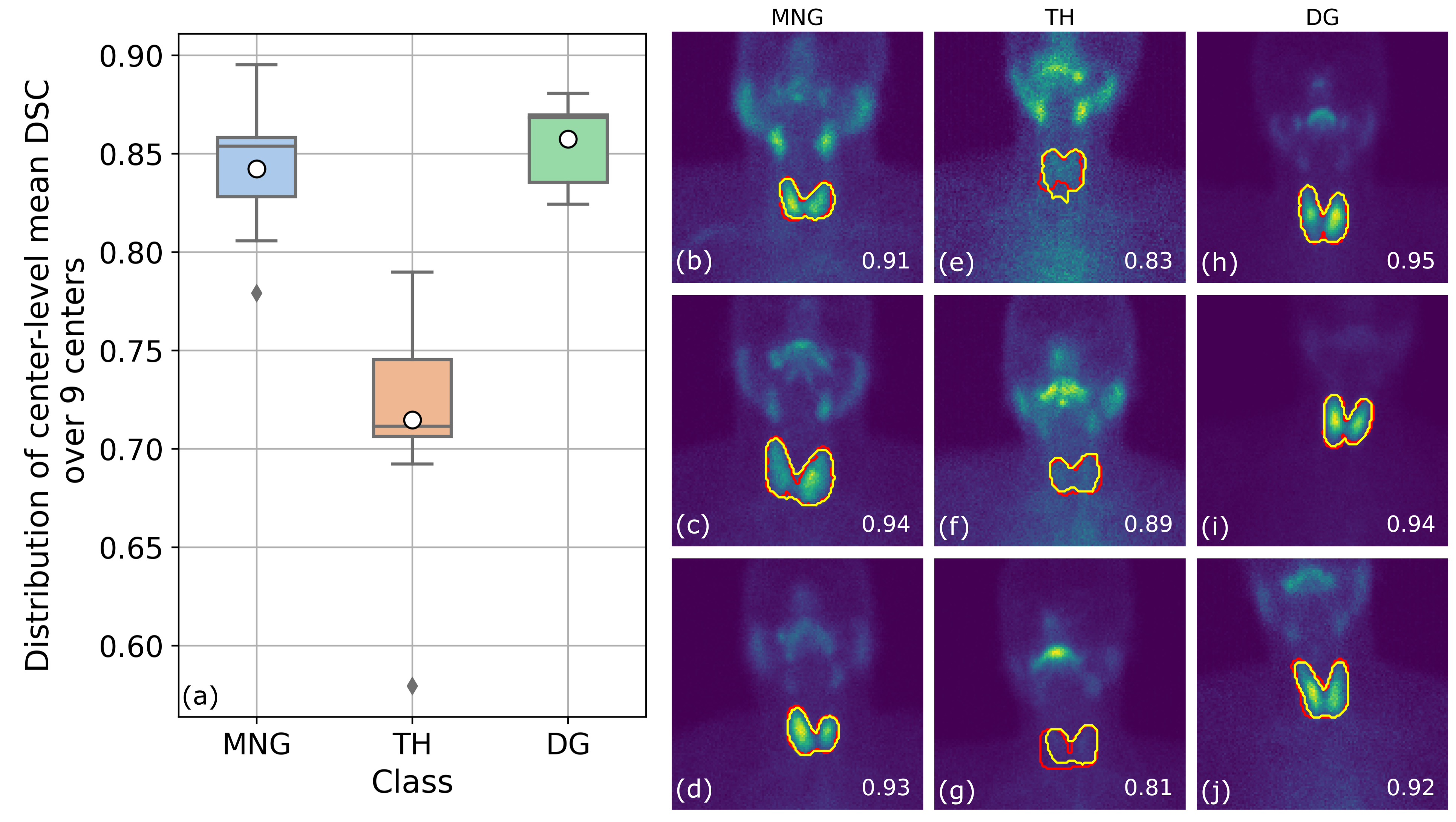}
\caption{(a) Distribution of center-level mean DSC over 9 centers for the classes, MNG, TH and DG. (b)-(d), (e)-(g), and (h)-(j) show some representative images from each class with the ground truth (red) and ResUNet predicted (yellow) segmentation of thyroid. The DSC between ground truth and predicted masks is shown in the bottom-right of each figure.}
\label{Fig:segmentation_plots}
\end{figure}

\begin{figure}[!ht]
\centering
\includegraphics[width=0.7\columnwidth]{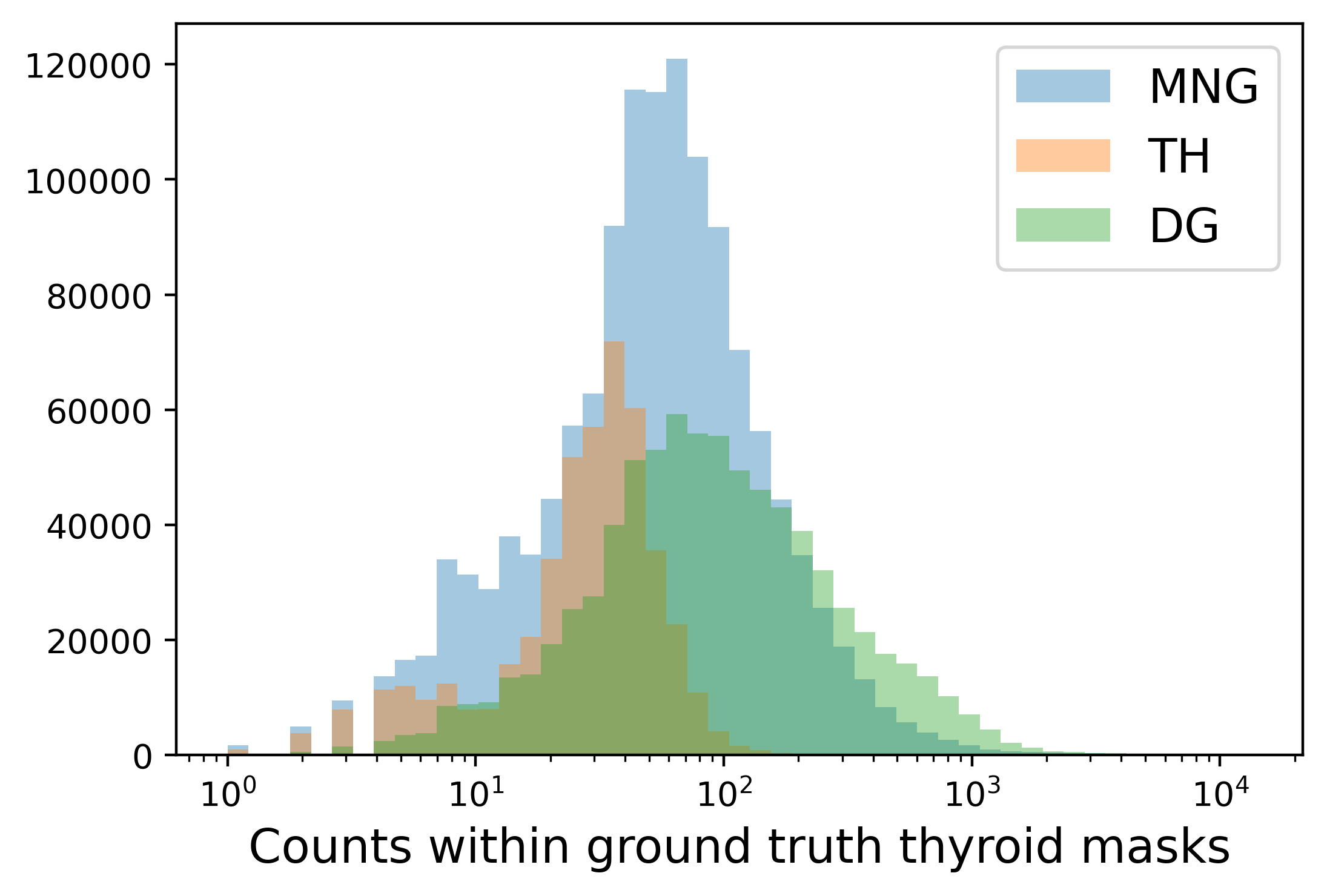}
\caption{Distribution of counts within the ground truth thyroid masks for the three thyroid pathology classes, MNG, TH and DG.}
\label{Fig:counts_within_thyroid}
\end{figure}

The nine trained segmentation models were tested under the LOCOCV scheme on their respective test centers. The distribution of DSC over different centers for each of the pathology class has been shown in Fig. \ref{Fig:segmentation_plots} (a), while some representative images comparing the ground truth and predicted masks have been shown in Fig. \ref{Fig:segmentation_plots} (b)-(j). For the classes MNG, TH and DG, the average of the mean DSC over 9 centers were 0.84$\pm$0.03, 0.71$\pm$0.06, and 0.86$\pm$0.02, respectively. Fig. \ref{Fig:counts_within_thyroid} shows the distribution of the number of counts within the ground truth segmented thyroid ROIs. Since the images from classes MNG and DG contain higher distribution of counts within thyroid than the TH class, this was reflected in their segmentation performance as well, with the MNG (counts range $\approx [1, 5 \times 10^3$]) and DG (counts range $\approx [1,1.3\times 10^4]$) classes being easier to segment by the network (counts in range) and TH (counts range $\approx [1,600]$) class being the hardest to segment. This finding is in agreement with the study \cite{resunet}, which shows that segmentation-based networks are generally biased towards segmenting regions with higher intensities or activities.

\subsection{Classification using features extracted via ResUNet segmentation}
In this scenario, as shown in Fig. \ref{Fig: Metrics} (d), the micro-average method yielded a mean accuracy of 0.74$\pm$0.05, ROC AUC of 0.90$\pm$0.02, and PRC AUC of 0.81$\pm$0.04 across all centers. The macro-averaging showed a mean precision of 0.76$\pm$0.04, recall of 0.77$\pm$0.05, F1-score of 0.76$\pm$0.05, and ROC AUC of 0.89$\pm$0.03. The weighted-average method had a mean precision of 0.77$\pm$0.06, recall of 0.74$\pm$0.05, F1-score of 0.74$\pm$0.05, and ROC AUC of 0.88$\pm$0.04.

We also analyzed the results for each class individually, as illustrated in Fig. \ref{Fig: Metrics} (b). The TH class again indicated best performance with a mean precision of 0.89$\pm$0.05, recall of 0.90$\pm$0.05, F1-score of 0.90$\pm$0.04, ROC AUC of 0.98$\pm$0.02, and PRC AUC of 0.95$\pm$0.05 across all centers. The MNG class obtained a mean precision of 0.70$\pm$0.16, recall of 0.68$\pm$0.13, F1-score of 0.67$\pm$0.10, ROC AUC of 0.83$\pm$0.05, and PRC AUC of 0.75$\pm$0.13. The DG class reached a Precision of 0.70$\pm$0.14, Recall of 0.73$\pm$0.10, F1-score of 0.70$\pm$0.07, ROC AUC of 0.86$\pm$0.04, and PRC AUC of 0.73$\pm$0.11. 

We compared the results of classification based on physician delineated and ResUNet predicted ROIs by computing the mean and standard deviation of their difference. The former obtained marginally higher results in all averaging metrics with a mean difference of 0.021$\pm$0.004, with the highest difference equal to 0.032 for micro-averaged PRC AUC. Similarly, we compared the two sets of results for each class indivaidually. The former again achieved marginally better results with a mean difference of 0.023$\pm$0.013 and the highest difference equal to 0.054 for PRC AUC of the MNG class. Although the results are only marginally higher in the first scenario, ResUNet was able to segment the thyroid and achieve a reasonable accuracy (mean DSC of 0.83$\pm$0.13 over all images from all centers).

We performed two one-sided paired $t$-test (TOST) at a significance level $\alpha = 5\%$ to access the equivalence of the classification metrics obtained from scenarios 1 and 2, under the null hypothesis that the two metrics are not equivalent. For this testing, we utilized five class-wise metrics: precision, recall, F1-score, ROC AUC, and PRC AUC over the nine centers. For MNG class, the values of precision and F1-score were equivalent with TOST $p$-values of 0.04 each respectively, while recall, ROC AUC, and PRC AUC were not with $p$-values of 0.18, 0.13 and 0.56, respectively. Similarly, for TH class, only ROC AUC and PRC AUC were equivalent with $p$-values of $2.4\times10^{-4}$ and 0.03, respectively, while precision, recall and F1-score were not, with $p$-values of 0.25, 0.16, and 0.17, respectively. Finally, for DG class, precision, recall, F1-score and ROC AUC were equivalent with $p$-values of $5.2\times10^{-3}$, 0.05, $3.0\times10^{-3}$, and 0.01, respectively, while PRC AUC was not with a $p$-value of 0.14. 

This shows that out of the 5 classification metrics compared (between scenarios 1 and 2) over three classes, 8 out of 15 metrics were statistically equivalent (while the rest being very close to each other for the two scenarios). This shows that on an average, the classification method based on extracting features from physician delineated ROIs is equivalent to the classification based on our fully-automated pipeline which utilized ResUNet predicted ROIs for feature extraction.

The offered automatic pipeline can reduce the time required for manual delineation of ROIs, speed up diagnostic processes, and ensure reproducibility which can lead to minimizing inter- and intra-observer variability. It can also serve as an assistant to physicians, particularly less experienced ones, by allowing them to compare their diagnostic results with those of the model. If the results match, it would provide validation; if they differ, it would prompt further consideration and possibly a second opinion from another physician.

In future work, we plan to expand the variability of the dataset to further enhance the pipeline's generalizability. Additionally, we aim to include multimodal data such as demographic data, clinical reports, and anatomical information from thyroid ultrasound images — essentially all the information a physician uses for diagnosis — alongside scintigraphy data to improve the pipeline's performance. However, it is important to note that even without these additional data, the results from features from physician delineations and the fully-automated pipeline were comparable over several metrics across all classes, underscoring the value and significance of scintigraphy images for classifying thyroid pathologies.

Although this study featured a diverse dataset from nine different centers encompassing various vendors, it is limited to data from a single country. Incorporating more diverse data from around the world could further enhance the generalizability of the pipeline. Additionally, the absence of detailed information such as the exact amount of radiopharmaceutical injection, the precise distance of the patient from the detector (zooming), the time of acquisition after injection, and the acquisition time may limit the model’s performance. Including these details could provide the model with more insights, potentially improving the pipeline's accuracy. Moreover, since the classification model was trained on features extracted from physician segmentations, we expected our classification results to be highly dependent on expert segmentations. For this study, we conducted all investigations with this limitation in mind and with a goal to produce a proof-of-concept and a functional model for thyroid disease classification. In future work, we aim to employ methods that perform segmentation-free classification of thyroid disease from scintigraphy images \cite{toosi2024segmentation}. 

\section{Conclusion}
This study effectively demonstrates that the automated pipeline, utilizing ResUNet for segmentation, achieves promising performance in delineating thyroid ROIs from scintigraphy images as well as delivers comparable results in pathology classification over several metrics to the scenario in which ROIs were delineated by physicians. These findings underscore the potential of AI-assisted tools to enhance diagnostic efficiency and consistency in the routine assessment of thyroid pathologies.

\section*{Acknowledgment}
This work was supported by the Natural Sciences and Engineering Research Council of Canada (NSERC) Discovery Grant RGPIN-2019-06467, as well as the BC Cancer Foundation. 

\section*{Co-authors contribution}
M. Sabouri and S. Ahamed designed the experiments, performed data analysis, developed models, and wrote the manuscript. A. Asadzadeh provided expert annotations. A. H. Avval performed literature review. S. Bagheri, M. Arabi, S. R. Zakavi, E. Askari, A. Rasouli, A. Aghaee, M. Sehati were responsible for providing data. G. Hajianfar contributed to technical discussion. F. Yousefirizi, C. Uribe, G. Hajianfar, H. Zaidi, and A. Rahmim discussed research efforts and provided feedback on the manuscript.

{  
\bibliographystyle{IEEEtran}
\bibliography{References}  
}  

\end{document}